\documentclass[usenatbib,fleqn]{mn2e}

\usepackage{graphicx}
\usepackage{float}
\usepackage{epstopdf}
\usepackage{amsmath}
\usepackage{amssymb}
\usepackage[labelformat=empty,labelsep=none]{subcaption}
\usepackage{epsfig}
\usepackage{times}
\usepackage{multirow}
\usepackage{color}
\usepackage{threeparttable}

\def\gtsim {\gtrsim}   % Greater than sim
\def\ltsim {\lesssim}   % Less than sim
\def\fo {f_{\rm O}}

\newcommand{\dummytitle}[1]{}

\newcommand{\msun}{{\,\rm M}_\odot}

\title[Oxygen loss from galaxies]{Oxygen Loss from Simulated Galaxies and the Metal Flow Main Sequence: Predicting the Dependence on Mass and Environment}%[] short title that appears at top of pages,{} actual title
\author[P.~Taylor, C.~Kobayashi, and L.J.~Kewley]{Philip~Taylor$^{1,2}$\thanks{E-mail: philip.1.taylor@anu.edu.au}, Chiaki~Kobayashi$^{3,2}$, and Lisa~J.~Kewley$^{1,2}$\\
$^1$Research School of Astronomy and Astrophysics, Australian National University, Canberra, ACT 2611, Australia\\
$^2$ARC Centre of Excellence for All Sky Astrophysics in 3 Dimensions (ASTRO 3D), Australia\\
$^3$Centre for Astrophysics Research, School of Physics, Astronomy and Mathematics, University of Hertfordshire, Hatfield, AL10 9AB, UK}

\begin{document}

\date{Accepted  Received ; in original form}

\pagerange{\pageref{firstpage}--\pageref{lastpage}} \pubyear{}

\maketitle

\label{firstpage}

%%%%%%%%%%%%%%    abstract    %%%%%%%%%%%%%%%%%
\begin{abstract}
We predict the mass fraction of oxygen lost from galaxies in a cosmological simulation as a function of stellar mass and environment at the present day.
The distribution with stellar mass is bimodal, separating star-forming and quenched galaxies.
The metallicity of gas and stars is self-consistently calculated using a chemical evolution model that includes supernovae type II and Ia, hypernovae, and asymptotic giant branch stars. 
The mass of oxygen lost from each galaxy is calculated by comparing the existing oxygen in gas and stars in the galaxy to the oxygen that should have been produced by the present-day population of stars. 
More massive galaxies are able to retain a greater fraction of their metals ($\sim 100$ per cent) than low-mass galaxies ($\sim 40 - 70$ per cent).
As in the star formation main sequence, star-forming galaxies follow a tight relationship also in terms of oxygen mass lost -- a metal flow main sequence, ZFMS -- whereas massive quenched galaxies tend to have lost a greater fraction of oxygen (up to 20 per cent), due to AGN-driven winds.
The amount of oxygen lost by satellite galaxies depends on the details of their interaction history, and those in richer groups tend to have lost a greater fraction of their oxygen.
Observational estimates of metal retention in galaxies will provide a strong constraint on models of galaxy evolution.

\end{abstract}


%%%%%%%%%% introduction %%%%%%%%%%%%%%%%%%%%%%

\section{Introduction}
\label{sec:intro}

%Galaxy evolution results from the complex interplay of gravity, hydrodynamics, and thermodynamics acting on scales smaller than a parsec to larger than a megaparsec.
The baryon cycle of a galaxy plays a large role in determining its properties.
Molecular gas fuels star formation, and accretion of gas onto the central supermassive black hole (BH) can lead to galaxy-wide feedback and suppression of star formation.
Stars, in turn, process hydrogen and helium into heavier elements (metals), returning them to the inter-stellar medium (ISM) if the stars explode as a supernova.
The energy injected into the ISM from stellar or active galactic nucleus (AGN) feedback can drive enriched gas outflows into the circumgalactic {and intergalactic media} \citep[CGM, IGM; e.g.,][]{lanzetta95,cowie98,sargent98,pettini01,ohyama02,kraft09,feruglio10,cicone12,tumlinson17}, and low metallicity {or primordial} gas can be accreted from the cosmic web, diluting the metallicity of the ISM.
Therefore, the metal content of a galaxy and the surrounding CGM and IGM can give information on its evolutionary history.

Gas and stellar metallicities are derived from emission- and absorption-line spectra.
Linear combinations of stellar template spectra convolved with a line-of-sight velocity distribution are fitted to the observed spectrum, allowing single stellar population (SSP) parameters, such as age, metallicity, and [$\alpha$/Fe] to be inferred {\citep{worthey94,vazdekis10,maraston11,conroy13}.}
Subtracting the fitted spectrum from the observed spectrum leaves emission lines from ionised gas; these are typically fit with one or more Gaussian functions to obtain the flux in each line.
Combinations of ratios of these fluxes are used to derive the gas-phase metallicity, though there can be significant differences in the absolute values of metallicity obtained depending on the calibration used \citep{kewley08,kewley19}.
The metallicity of the CGM is measured from absorption features in the spectra of quasars on sight lines that pass though the CGM \citep[e.g.,][]{tumlinson11}.

The total mass of metals ever produced in a galaxy must be known in order to estimate the fraction of metals that have been lost.
For this reason, most studies to date have focused on nearby dwarf galaxies in which individual stars can be resolved to construct a colour--magnitude diagram (CMD), from which the star formation history (SFH) of the galaxy can be derived \citep[e.g.,][]{mcquinn15cmd}.
By assuming a constant fraction of metals produced per mass of stars, the total mass of metals produced is obtained by integrating over the SFH.
Oxygen is often used as a proxy for total metallicity for a number of reasons: emission lines from ionised oxygen are strong and readily available in optical wavelengths, and are sensitive to the electron temperature, which is related to the total metal content; oxygen is the most abundant metal; and it is produced almost completely by core-collapse supernovae (i.e., type II supernovae and hypernovae) on short timescales, so the instantaneous recycling approximation is valid.

Local dwarf galaxies are found to have lost 95 to 99 per cent of produced oxygen, due primarily to supernova feedback \citep[e.g.,][]{kirby11,mcquinn15}.
\citet{telford19} analysed the third of M31 covered by the Panchromatic Hubble Andromeda Treasury \citep[PHAT,][]{dalcanton12}, finding that $\sim62$ per cent of all metal mass has been lost from the central 19 kpc of the galaxy.
In more distant galaxies, direct measurements are more challenging since individual stars cannot be resolved, however \citet{belfiore16} obtained spatially resolved measurements of oxygen loss in the face-on spiral galaxy NGC 628, with about half of oxygen mass lost from the inner 7 kpc.

{Galaxies become `quenched' when they are no longer able to form stars, either because they have depleted their reservoir of cold, dense gas \citep[e.g.,][]{quai19}, or because their gas has been heated and/or removed by stellar and/or AGN feedback \citep[e.g.,][ see also \citealt{bower17,pt17a} for theoretical studies]{lynds67,heckman00,pettini00,ohyama02,kraft09,feruglio10,cicone12,tombesi13,teng14}.
These feedback processes redistribute gas and metals with the galaxy, and galaxy mergers can also greatly affect stellar orbits.
Furthermore, inflows of pristine or pre-enriched gas from outside the galaxy can dilute existing enriched material.
Therefore the spatial distribution of both stellar and gas-phase metals contains a wealth of information about a galaxy's evolutionary history \citep[e.g.,][]{ck04,pt17b,jianhui18,torrey18,zinchenko19,torrey19,pt19a,belfiore19,trayford19,laralopez19}.}

Theoretically, hydrodynamic simulations of galaxy formation and evolution are able to track the stellar and chemical evolution that leads to gas enrichment, as well as the processes that redistribute gas inside and outside galaxies.
To date, few works have predicted the metal loss from simulated galaxies.
{\citet{ck07} (their Fig. 16) showed a dependence of metal loss fraction on galaxy total mass in a cosmological simulation.}
{Such a dependence was also seen in the cosmological simulations of \citet{barai11}, and in fifteen galaxies from the Feedback In Realistic Environments \citep[FIRE,][]{hopkins14fire} simulations studied by \citet{ma16}}.
Using the cosmological simulation of \citet{pt15a}, \citet{pt15b} characterised AGN-driven, metal-enriched outflows, noting a significant reduction in the metal content of the host galaxy, but did not investigate the overall fraction of metals lost by the present day.
{More recently, \citet{nelson19} demonstrated the efficacy of stellar- and AGN-driven outflows at enriching the IGM in the TNG50 simulation as a function of both galaxy stellar mass and redshift.}

In this paper, we present results from our cosmological, hydrodynamical simulation that includes a detailed prescription for chemical evolution.
We show the dependence on mass of a galaxy's ability to retain oxygen, and, for the first time, predict the dependence on environment.
Section \ref{sec:sims} briefly describes the details of the simulation.
In Sections \ref{sec:results} and \ref{sec:haloes} we present our results for both galaxy and halo scales, and compare our predictions to observations and other models in Section \ref{sec:theorycomp}.
Finally, in Section \ref{sec:discussion}, we discuss our results in a wider context and summarise our conclusions.

\section{The Simulation}
\label{sec:sims}

\begin{figure*}
	\centering
	\includegraphics[width=\textwidth,keepaspectratio]{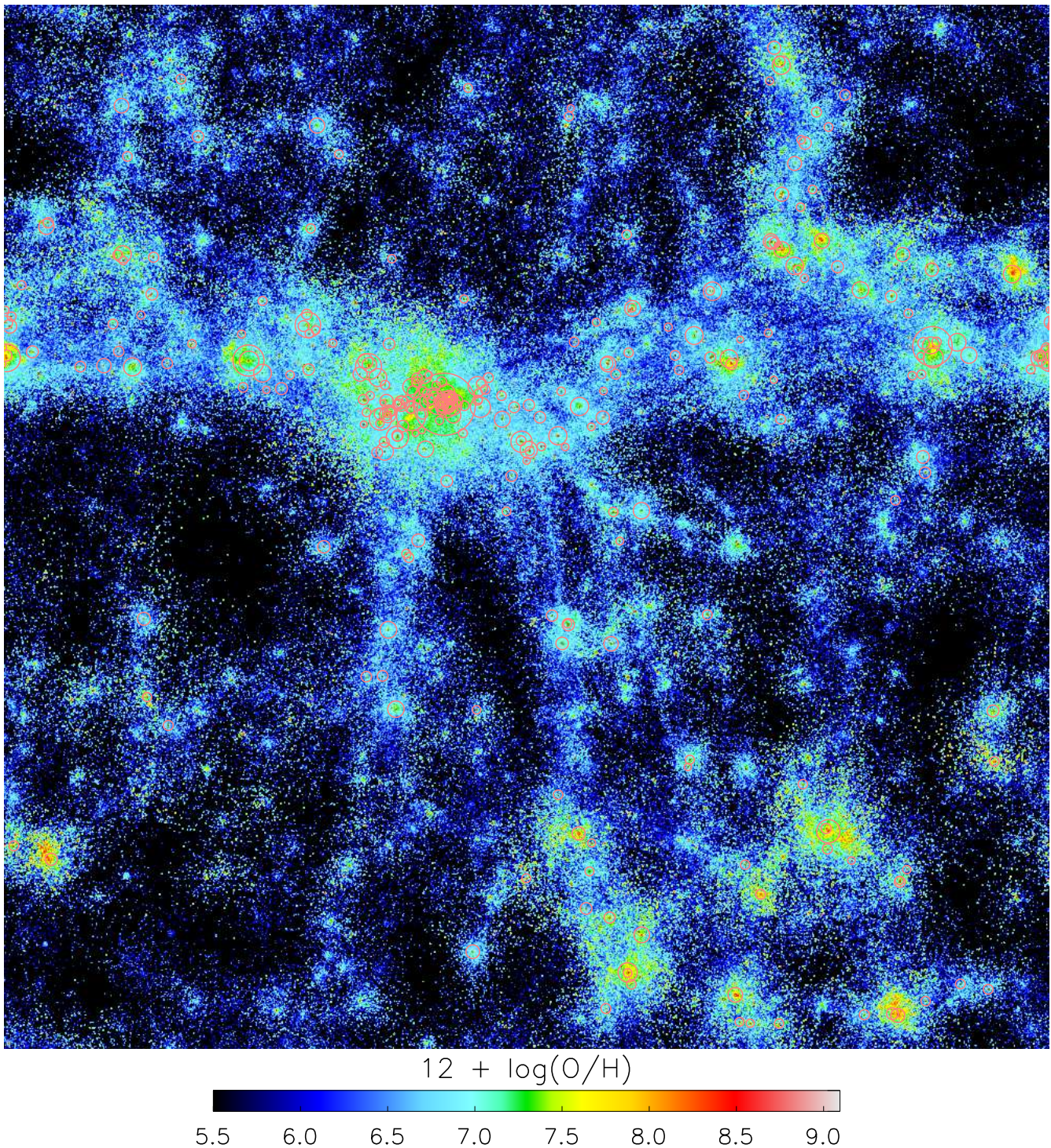}
	\caption{Spatial distribution of gas-phase oxygen in our full simulation box.
	Solar oxygen abundance is $12+\log({\rm O}/{\rm H}) = 8.78$.
	Salmon-coloured circles show $R_{200}$ for all galaxies with stellar mass $M_* > 10^{10}\msun$.}
	\label{fig:fullbox}
\end{figure*}

The simulation analysed in this paper is a cosmological, chemodynamical simulation, introduced in \citet{pt15a}, {which we have previously shown reproduces the observed $M_*$ -- SFR and $M_*$ -- metallicity relations, among others (Figs. 6, 8, and 12, respectively, of \citealt{pt16}).
The distribution of gas-phase oxygen in our full simulation box is shown in Fig. \ref{fig:fullbox}.}
Our simulation code is based on the SPH code {\sc gadget-3} \citep{springel05gadget}, updated to include:  star formation, energy feedback and chemical enrichment \citep{ck04,ck07} from supernovae \citep[SNe II and Ia,][]{ck09} and hypernovae \citep{ck06,ck11a}, and asymptotic giant branch (AGB) stars \citep{ck11b}; heating from a uniform, evolving UV background \citep{haardt96}; metallicity-dependent radiative gas cooling \citep{sutherland93}; and a model for BH formation, growth, and feedback \citep{pt14}, described in more detail below.
We use the IMF of stars from \citet{kroupa08} in the range $0.01-120\msun$, with an upper mass limit for core-collapse supernovae of $50\msun$.
{Stars can form from any gas particle that is converging, cooling, and Jeans unstable \citep{katz92}, and has not experienced heating from stellar or AGN feedback in the current timestep.
Note that we do not use an explicit density threshold as in other works \citep[e.g.,][]{vogelsberger13,dubois14,schaye15,tremmel17}, though both the cooling and dynamical timescales do depend on gas density.}

{For the chemical evolution, we use the self-consistent nucleosynthesis yields from \citet{ck11b}, which match the observed evolution of elemental abundances from C to Zn (except for Ti).
With the adopted nucleosynthesis yields, the IMF-weighted yields are 0.021 and 0.019 at $Z=0$ and $Z_\odot$, respectively, and the net yields, defined as $y/(1-R)$, where $R$ denotes the return fraction, are 0.037 and 0.039, respectively.
We do not assume the instantaneous recycling approximation, and the stellar mass loss is also calculated self-consistently.}

{As in our previous works, the} initial conditions consist of $240^3$ particles of each of gas and dark matter in a periodic, cubic box $25\,h^{-1}$ Mpc on a side, giving spatial and mass resolutions of $1.125\,h^{-1}$ kpc and $M_{\rm DM}=7.3\times10^7\,h^{-1}\msun$, $M_{\rm g}=1.4\times10^7\,h^{-1}\msun$, respectively.
We employ a WMAP-9 $\Lambda$CDM cosmology \citep{wmap9} with $h=0.7$, $\Omega_{\rm m}=0.28$, $\Omega_\Lambda=0.72$, $\Omega_{\rm b}=0.046$, and $\sigma_8=0.82$.

{In our model} BHs form from gas particles that are metal-free and denser than a specified critical density, mimicking the most likely formation channels in the early Universe via direct collapse of a massive gas cloud \citep[e.g.,][]{bromm03,koushiappas04,agarwal12,becerra15,regan16a,hosokawa16} or as the remnant of Population {\sc III} stars \citep[e.g.,][]{madau01,bromm02,schneider02}.
The BHs grow through gas accretion and mergers.
The accretion rate is estimated assuming Eddington-limited Bondi-Hoyle accretion {\citep{bondi44}}:
\begin{equation}\label{eq:macc}
	\dot{M}_{\rm acc} = \frac{4\pi G^2 M^2_{\rm BH}\rho}{\left(c^2_s + v^2\right)^{3/2}},
\end{equation}
where $M_{\rm BH}$ is the mass of the BH, $\rho$ the local gas density, $c_s$ the local sound speed, and $v$ the speed of the BH relative to the local gas particles.
Two BHs merge if their separation is less than the gravitational softening used, and their relative speed is less than the local sound speed.
A fraction of the energy liberated by gas accretion is coupled to neighbouring gas particles in a purely thermal form.
{Although our seeding method is different from the merger-driven model, the resulting BH masses are $\sim10^5\msun$ when the dark matter haloes become $10^{10}\msun$ \citep[for more details, see][]{wang19}}.

%%%%%%%%%%%%%%%% res %%%%%%%%%%%%%%%%%%%%%%%

\section{Oxygen Loss from Galaxies}
\label{sec:results}

Galaxies are identified using a Friends-of-Friends (FoF) algorithm.
The code associates dark matter particles, separated by at most 0.02 times the mean inter-particle separation, into groups.
Gas, star, and BH particles are then joined to the group of their nearest DM neighbour.
The virial radius of each galaxy is estimated as the radius of a spherical region centred on the galaxy whose mean density is 200 times the critical density of the Universe, $R_{200}$.
The total stellar mass ($M_*$) and star formation rate (SFR) of each FoF galaxy determine its position relative to the star-forming main sequence \citep[SFMS;][]{wuyts11,renzini15}.
In \citet{pt17b} we identified as star forming galaxies those whose distance from the SFMS, $\Delta$SFMS, satisfied $\Delta{\rm SFMS} > -0.5$\,dex.

We estimate the retention/loss fraction of oxygen in similar manner to observers \citep[e.g.,][]{mcquinn15}.
The oxygen mass associated with each gas and star particle can be read directly from the simulation output.
For each star particle, we calculate its net yield of oxygen, $y(A,Z_*)$, given its age $A$, and metallicity $Z_*$\footnote{The initial $Z_*$ is the same as the present $Z_*$ because internal mixing is not included.}.
The fraction of oxygen produced by the present-day population of stars that has been lost from the galaxy is then
\begin{equation}\label{eq:fo}
	1-\fo = \frac{\sum y(A,Z_*)M_{*,{\rm init}} - \left(\sum M_{*,{\rm O}} + \sum M_{\rm g,O}\right)}{\sum y(A,Z_*)M_{*,{\rm init}}},
\end{equation}
where $M_{*,{\rm init}}$ is the initial mass of the star particle (which is larger than the current mass of the particle due to stellar mass loss), and $M_{*,{\rm O}}$ and $M_{\rm g,O}$ are the current mass of oxygen within a star or gas particle, respectively.%, and the sums run over all particles in the galaxy.

\begin{figure}
	\centering
	\includegraphics[width=0.48\textwidth,keepaspectratio]{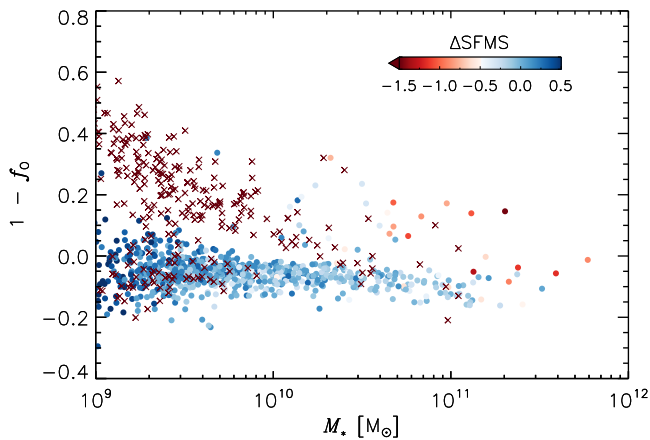}
	\caption{Mass fraction of oxygen lost from galaxies at $z=0$ as a function of stellar mass.
	Points are coloured by distance from the SFMS, and crosses denote galaxies with no star formation in the last $10^8$ years.}
	\label{fig:foms}
\end{figure}
We show $1-\fo$ for the FoF galaxies as a function of galaxy stellar mass at $z=0$ in Fig. \ref{fig:foms}.
Points are coloured according to $\Delta$SFMS, with blue points belonging to the star forming main sequence, red being quenched galaxies, and crosses denoting galaxies with no stars younger than $10^8$ years.
The galaxy population is bimodal; star-forming galaxies mostly occupy a tight sequence (which we refer to as the metal flow main sequence, or ZFMS) with $-0.2 < 1-\fo < 0$, whereas quenched galaxies tend to have $1-\fo > 0$.
Note that $1-\fo < 0$ implies that these galaxies contain more oxygen than their stars could have produced; this is addressed below. 
The distribution of quenched galaxies shows a strong mass dependence whereby low mass galaxies tend to have lost a larger fraction of their oxygen (larger $1-\fo$) than more massive galaxies.
In these galaxies, stellar and AGN feedback \citep[depending on the galaxy mass, e.g.,][]{pt17a} heat and remove enriched gas from the galaxy.
This trend is very similar to \citet{ck07}, who used a different analysis method, and is explained due to the deeper potential well of massive galaxies.
%There is also a population of quenched galaxies in the same region as the star-forming galaxies, particularly at $M_*<10^{10}\msun$; these are satellite galaxies that have undergone tidal stripping of gas and stars by a more massive companion, and are discussed further below.

To further understand the distribution of $1-\fo$, we split the FoF galaxies into centrals, satellites, and isolated by the following method.
For the most massive galaxy not yet labelled, we find its $R_{200}$ and identify the $N_{\rm group}$ galaxies within this distance, including itself.
If $N_{\rm group}$ is 1, this galaxy is isolated; otherwise it is a central and the galaxies identified near it are labelled satellites.
{Note that numerous different measures of galaxy environment are used in the literature, typically using $n^{\rm th}$-nearest neighbour distance \citep[e.g.,][]{pt17b}, or the number of galaxies within a fixed aperture to obtain a quantitative estimate of environmental density.
All environmental metrics have their own strengths and drawbacks \citep[e.g.,][]{haas12,muldrew12}; we use the discrete splitting into centrals, satellites, and isolated since this allows for a relatively straightforward understanding of the results presented below.}

\begin{figure*}
	\centering
	\includegraphics[width=\textwidth,keepaspectratio]{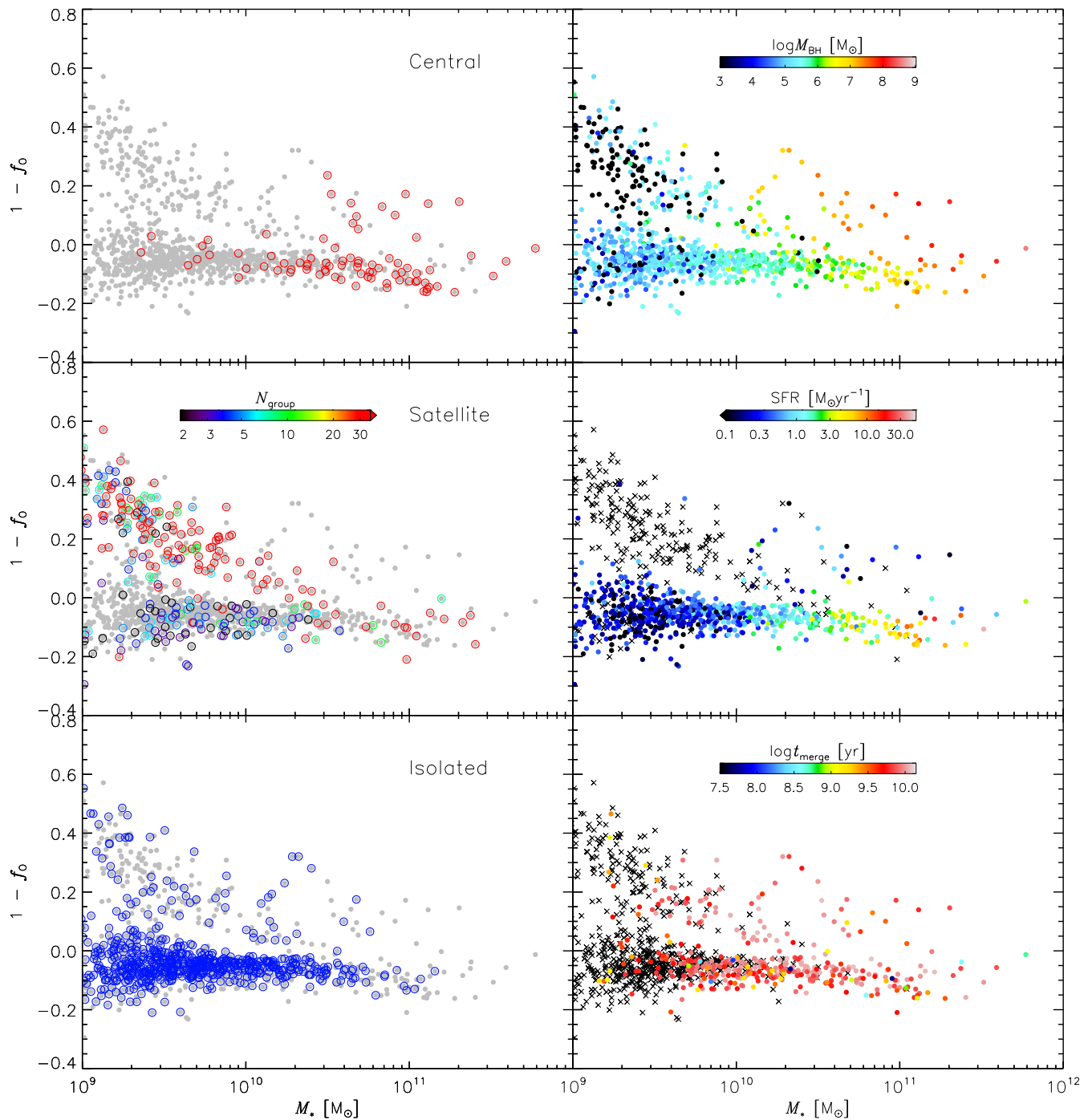}
	\caption{Mass fraction of oxygen lost from galaxies at $z=0$ as a function of stellar mass.
	\emph{Left column}: galaxies are separated into central (top panel), satellite (middle panel), and isolated (bottom panel), as defined in the text.
	Satellite galaxies are coloured by the number of members of the group to which they belong.
	\emph{Right column}: points are coloured by $M_{\rm BH}$ (top panel), SFR averaged over $10^8$ years (middle panel), and time since last major merger (stellar mass ratio larger than 1/4; bottom panel).
	Black crosses denote galaxies with no star formation in the last $10^8$ years (middle panel), or that never experienced a major merger (bottom panel).}
	\label{fig:env}
\end{figure*}

In Fig. \ref{fig:env} we again show $1-\fo$ as a function of stellar mass, but now separate the population based on both environment (left column) and properties indicative of internal processes, specifically $M_{\rm BH}$ (top panel), SFR (middle panel), and time since the most recent major merger, $t_{\rm merge}$ (bottom panel)\footnote{We define a major merger as one with a stellar mass ratio of at least 1/4.}.
In each panel in the left column the grey points show the full population of galaxies, and the coloured circles show the sub-population of interest.
The central galaxies (top panel) tend to be the most massive, and are found along and above the ZFMS.
Above $\sim 3\times10^{10}\msun$, most of the galaxies above the main sequence are centrals, and correspond to the quenched galaxies in Fig. \ref{fig:foms}.
It is at this mass that AGN feedback can start to effectively quench star formation by heating and removing gas from galaxies \citep{kauffmann03,baldry06,bower17,pt17a} if sufficient gas accretes onto the BH.
Furthermore, we showed in \citet{pt15b} that such AGN-driven winds eject metals into the CGM and IGM.
{It is also clear from the top right panel that these galaxies host BHs that are significantly more massive than galaxies of the same stellar mass that lie on the ZFMS, further suggesting the role of AGN feedback.}

The left middle panel of Fig. \ref{fig:env} highlights satellite galaxies, with colours corresponding to the richness of the groups.
Satellite galaxies are found on both the ZFMS and the quenched sequence, with quenched satellites tending to reside in richer groups {(galaxies with no star formation in the last $10^8$ years are shown as black crosses in the middle right panel)}.
This is clear evidence for environmental quenching in the simulation, whereby satellite galaxies falling into a rich cluster are quenched due to their interactions with the host galaxy, other satellites, and the intra-cluster medium (ICM) itself \citep[e.g.,][]{tal14,schaefer17,schaefer19}.

The bottom left panel of Fig. \ref{fig:env} shows isolated galaxies; those with no companions within $R_{200}$.
These galaxies mostly occupy the ZFMS, though some are quenched, especially at lower masses.
The quenched, isolated galaxies with $M_*>10^{10}\msun$ have lost gas to large-scale outflows in their history; in one instance the outflow was driven by AGN feedback, in the other galaxies it was less clear if stellar or AGN feedback (or both) dominated.
These galaxies also all experienced major mergers in their past, which may have helped to fuel whichever feedback mechanism led to the gas outflows and associated loss of metals.
{The bottom right panel of Fig. \ref{fig:env} shows $1-\fo$ as a function of stellar mass and coloured by $t_{\rm merge}$.
There is no trend with $t_{\rm merge}$, and the majority of massive galaxies have experienced a major merger at some point in their history.
However, this figure does not show the details of individual mergers, the most important of which is the gas fraction of the merging galaxies \citep[see also Fig. 11 of][]{ck04}.
The detailed role of mergers in affecting $1-\fo$, and the production of starbursts, AGN activity, and outflows more generally, will be addressed in a future work.}

\section{Galaxy Haloes}
\label{sec:haloes}

\begin{table*}
\caption{Properties of the galaxies shown in Fig. \ref{fig:maps}.}
	\begin{threeparttable}
	\begin{tabular}[width=0.5\textwidth]{cccccccc}
		Galaxy & $\log M_*$ & SFR\tnote{a} & $R_{200}$ & Environment & Group Richness & $1-\fo$ & $1-\fo (r<R_{200})$ \\
		& $[\msun]$ & $[\msun\,{\rm yr}^{-1}]$ & $[{\rm kpc}]$ & & & &\\
		\hline 
		ga0008 & $1.63\times10^{11}$ & 22.3 & 342\tnote{b} & Satellite & 28 & -0.07 & 0.01 \\
		ga0012 & $1.30\times10^{11}$ & 0.3 & 399 & Central & 4 & 0.14 & 0.16 \\
		ga0135 & $1.37\times10^{10}$ & 2.4 & 212 & Isolated & 1 & -0.13 & -0.12 \\
		ga0569 & $7.61\times10^{9}$ & 0.0 & 72 & Isolated & 1 & 0.25 & 0.30 \\
		\hline
	\end{tabular}
	\begin{tablenotes}
		\item[a] SFR is averaged over the last $10^8$ years.
		\item[b] This value is calculated from equation \eqref{eq:r200} given the galaxy's stellar mass.
	\end{tablenotes}
	\end{threeparttable}
\label{tab:maps}
\end{table*}

In Section \ref{sec:results} we showed that many of our simulated galaxies -- star-forming galaxies in particular -- have $1-\fo<0$, indicating that they contain a greater mass of oxygen than can be accounted for by the evolution of their existing stellar populations.
These galaxies are defined by their FoF groups, which typically extend out to $\ltsim4-5$ times their effective radius, $R_{\rm e}$.
As such, they are good analogues for observed galaxies, but do not account for metal flows into and out of the galaxies.

\begin{figure}
	\centering
	\includegraphics[width=0.48\textwidth,keepaspectratio]{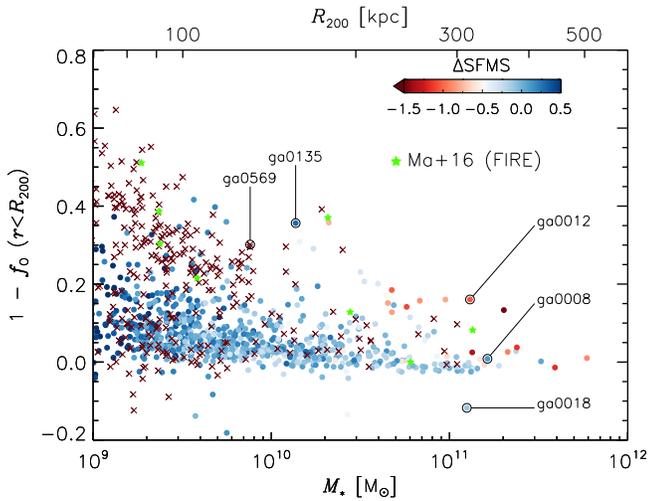}
	\caption{Same as Fig. \ref{fig:foms}, but we now consider all gas and stars within $R_{200}$ for each galaxy.
	{$R_{200}$, as given by equation \eqref{eq:r200}, is shown on the top axis.}
	Theoretical data from \citet{ma16} are included as the green stars.
	The four galaxies shown in Fig. \ref{fig:maps} are indicated.}
	\label{fig:R200SFMS}
\end{figure}

Fig. \ref{fig:R200SFMS} shows $1-\fo$ for our simulated galaxies, but we now consider all particles within $R_{200}$, with points coloured as in Fig. \ref{fig:foms}.
Note that some satellite galaxies are sufficiently embedded within the dark matter halo of their host that $R_{200}$ is not well defined.
We obtain a fit relating $R_{200}$ and $M_*$ for the central and isolated galaxies in our simulation, given by
\begin{equation} \label{eq:r200}
	\log R_{200} = -0.946 + 0.310\log M_*,
\end{equation}
and use this for all satellite galaxies to produce Fig. \ref{fig:R200SFMS}.
{This enables us to treat all of our simulated galaxies consistently.
Note, however, that since $R_{200}\propto M^{1/3}_{200}$ this implies a single power-law relationship between $M_*$ and $M_{200}$, whereas a broken power law is observed \citep[e.g.,][]{behroozi10,moster10}.
Due to the limited size of our simulation box, we have only a small number of halos more massive than the observed break in the power law, which may account for this discrepancy; see \citet{pt16} for more discussion.}

Most star-forming galaxies (blue points) have $1-\fo (r<R_{200}) \gtsim 0$, indicating that oxygen is either conserved or lost at galaxy halo scales into the IGM\footnote{Note that we do not track the inter-halo transport of oxygen in this figure, which could account for the scatter around $1-\fo=0$, especially at low mass.}.
A small number of galaxies still have $1-\fo (r<R_{200})\ltsim-0.05$.
We have explicitly checked the histories of galaxies with $1-\fo<-0.1$ (such as ga0018, indicated on Fig. \ref{fig:R200SFMS}) and find that such low values are transient, with the galaxies experiencing a close interaction with a neighbouring galaxy that makes it difficult for the FoF routine to associate star particles with one galaxy or the other.
In a future work, we will analyse the evolution of $1-\fo$ in detail for all our simulated galaxies.
We also show results from the Feedback In Realistic Environments simulations \citep[FIRE;][]{hopkins14fire} as the green stars \citep{ma16}.
The FIRE galaxies occupy the same region as our quenched galaxies, which may be due to the stronger stellar feedback in the FIRE model, despite their lack of AGN feedback.

\begin{figure}
	\centering
	\includegraphics[width=0.48\textwidth,keepaspectratio]{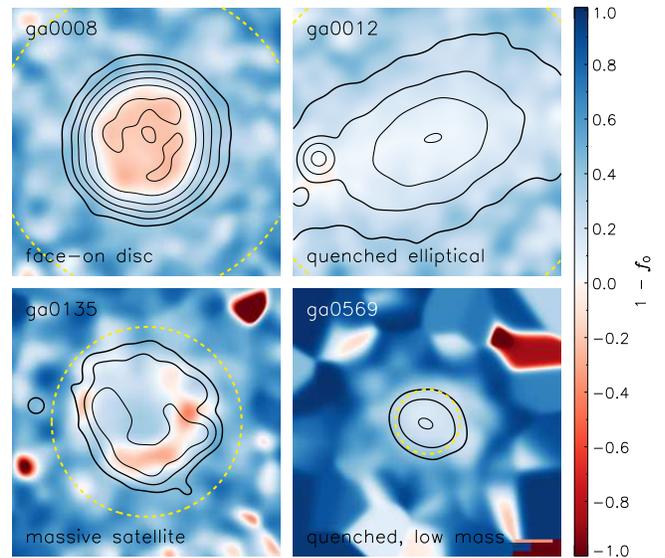}
	\caption{Maps of oxygen loss fraction, $1-\fo$, for four simulated galaxies.
	From top left to bottom right: ga0008 is a face-on disc; ga0012 experienced strong AGN-driven outflows; ga0135 is the third most massive member of a galaxy group; ga0569 is a low mass, quenched galaxy.
	Panels are $30$ kpc on a side.
	Contours show $V$-band surface brightness in the range 19 - 24 mag$\,$arcsec$^{-2}$, {and the dashed yellow circles show $R_{200}/20$}.}
	\label{fig:maps}
\end{figure}

Figs. \ref{fig:foms} and \ref{fig:R200SFMS} suggest that there must be a net transfer of $\sim 10$ per cent of the total oxygen produced from the outskirts of galaxies to their centres, with this effect particularly pronounced in star-forming galaxies.
To further investigate this, we show in Fig. \ref{fig:maps} maps of $1-\fo$ for a selection of galaxies with a variety masses, environments, and star-formation histories; {more properties of these galaxies are given in Table \ref{tab:maps}}.
Panels are $30$ kpc on a side and show (from top left to bottom right): a face-on disc galaxy (ga0008); a galaxy that has experienced strong AGN feedback and outflows in its past (ga0012); a star-forming galaxy with stellar mass close to $3\times10^{10}\msun$ (ga0135); and a low mass, quenched galaxy (ga0569).
The contours show $V$-band surface brightness from 19 (thinnest lines) to 24 mag$\,$arcsec$^{-2}$ (thickest lines). %, and some properties of the galaxies are given in Table \ref{tab:maps} \textcolor{red}{(ehh, not sure is needed).}.
In ga0008, a star-forming disk galaxy, the central regions have $1-\fo<0$, indicating a net gain of O, while the outer regions have a net loss of O.
This means that there has been an inflow of O-rich gas into the centre of this galaxy whereby gas is enriched by the stars in the outskirts of the galaxy and then cools and falls to the centre, adding to the metals already formed by the stars there.
It is unlikely that inter-halo transport plays a dominant role.
It is also possible that a more complex sequence of events takes place with gas cycling out of and in to the galaxy \citep{anglesalcazar17} leading to multiple enrichment phases for each gas particle; this cosmic baryon cycle will be examined in detail in a future work. %Note to self - would this leave chemical imprint?
Galaxy ga0135 is also star forming, but has a stellar mass close to the quenching transition mass of $3\times10^{10}\msun$ \citep{kauffmann03,baldry06,bower17,pt17a}.
In this galaxy, areas with $1-\fo<0$ are not centrally concentrated (as for ga0008), and roughly trace the brightest $V$-band isophote.
%This indicates that inside-out quenching is taking place in this galaxy.
{This is due to ongoing star formation in this region caused by a gas-rich minor merger.}
Galaxies ga0012 and ga0569 in Fig. \ref{fig:maps} are quenched (by AGN and stellar feedback, respectively), and do not show the central concentration of low $1-\fo$ seen in the star-forming galaxies because they have removed much of their central gas.
{These galaxies are likely to undergo inside-out quenching, while self-regulated star-forming disk galaxies may form inside-out and quench outside-in \citep{vincenzo20}.
The connection between inside-out/outside-in quenching and the $1-\fo$ map will be investigated further in a future work.}

\section{Comparison with Observational Work}
\label{sec:theorycomp}

\begin{figure}
	\centering
	\includegraphics[width=0.48\textwidth,keepaspectratio]{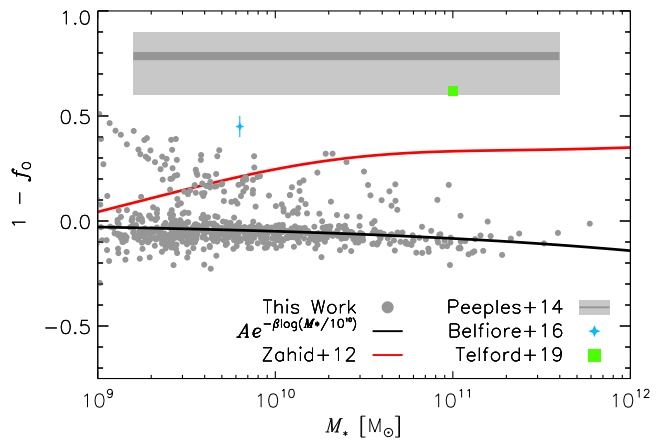}
	\caption{Mass fraction of oxygen lost from galaxies at $z=0$ as a function of stellar mass for star-forming galaxies only.
	Grey points show our simulated galaxies, the grey shaded region shows the prediction of the model of \citet{peeples14}, the red line shows a prediction of the model of \citet{zahid12}, the blue star shows the observational measurement of NGC 628 \citep{belfiore16}, and the green square is the observational measurement of M31 \citep{telford19}.}
	%\textcolor{red}{I am not sure about the usefulness of the fit to our data, and it is not (yet) discussed much in the text.}}
	\label{fig:comp}
\end{figure}

To date there are only a limited number of theoretical or observational works measuring the fraction of oxygen lost from galaxies.
Fig. \ref{fig:maps} demonstrates that the fraction of oxygen lost is not constant with radius (in qualitative agreement with \citet{belfiore16}), and aperture effects are important.
For NGC 628 and M31, oxygen loss fractions are given within 7 kpc and 19 kpc, respectively, whereas the model of \citet{peeples14} considers all material within 150 kpc of a galaxy.
\citet{belfiore16} also gave the radial dependence of oxygen retention; this can be readily obtained from hydrodynamical simulations, and may be a more useful metric against which to measure theoretical models, though this requires a greater sample of observational data.
In Fig. \ref{fig:comp} we show available data from the literature (in the stellar mass range relevant to this work) in comparison to our theoretical predictions for star-forming FoF galaxies (defined by $\Delta{\rm SMFS} > -0.5$).
The solid black line is a fit to our ZFMS of the form $1-\fo = Ae^{-\beta\log\left(M_*/10^{10}\msun\right)}$, with $A = -0.049\pm0.004$ and $\beta = -0.53\pm0.17$.
We use a bootstrap resampling technique whereby the parameters are re-derived for 5000 random selections of the data (in which some data are repeated, and some are absent), the same size as the original data set, to estimate the uncertainties as the widths of the resulting distributions.

The semi-empirical models of \citet{zahid12} and \citet{peeples14} are shown by the red line and grey shading, respectively.
For the \citet{zahid12} model, we use their recommended values for their free model parameters of $R=0.35$, $P_{\rm O}=0.007$, and $\Delta Z_{\rm g}=0$, denoting the fraction of gas returned from stars to the ISM, the fraction of oxygen produced per solar mass of gas that turns into stars (so that $y=P_{\rm O}/(1-R)$), and a constant offset applied to the gas metallicity, respectively.
For the redshift evolution of the gas-phase metallicity, we interpolate over the fits to observational data given in \citet{zahid14}.
We also show observationally derived values for NGC 628 \citep[blue star;][]{belfiore16} and M31 \citep[green square;][]{telford19}.

\begin{figure}
	\centering
	\includegraphics[width=0.48\textwidth,keepaspectratio]{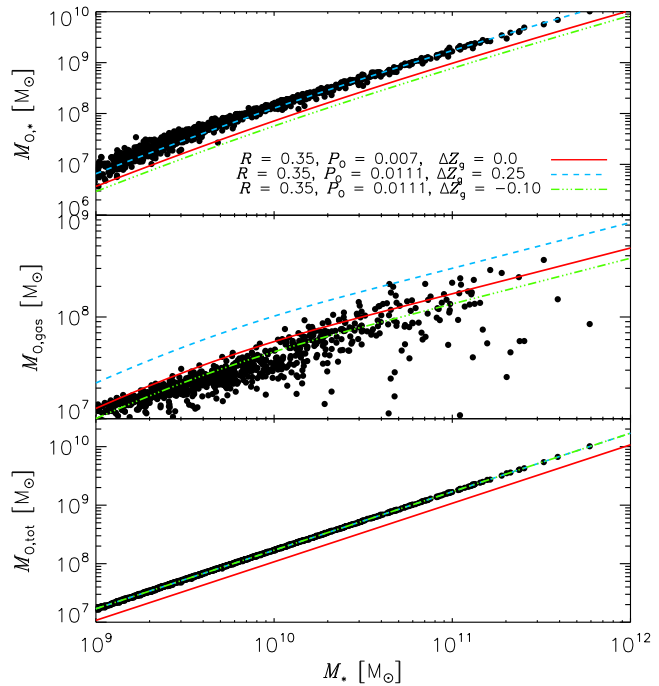}
	\caption{Predicted mass of oxygen locked up in stars (top panel), gas (middle panel), and the total mass of oxygen produced by stars, as a function of galaxy stellar mass.
	Galaxies from our simulation are shown by the black points, and lines show the model of \citet{zahid12} for a range of input parameters (see main text for details).}
	\label{fig:zahid12}
\end{figure}

It is clear that both the measurements for individual galaxies and the model region of \citet{peeples14} lie significantly above our simulated galaxies, indicating a larger fraction of oxygen has been lost than our model predicts.
Although the model of \citet{zahid12} makes more similar quantitative predictions as ours, the trend is opposite, with more massive galaxies losing a greater fraction of oxygen than less massive galaxies.

{In order to more completely compare to the model of \citet{zahid12}, we show in Fig. \ref{fig:zahid12} the mass of O locked in stars ($M_{\rm O,*}$; top panel), gas ($M_{\rm O,gas}$; middle panel), and the total O produced by the stars in a galaxy ($M_{\rm O,tot}$; bottom panel) as a function of stellar mass.
Our simulated galaxies are shown by the black points, and the \citet{zahid12} model (with different parameters) by the lines.
There is good qualitative agreement between the model and simulation, especially for $M_{\rm O,*}$ and  $M_{\rm O,tot}$, but their fiducial model (solid red line) is offset from the simulated data in each panel.
Although the oxygen yield is a function of age and metallicity for each star particle in our simulation, averaged over a galaxy it is well approximated by a constant, as shown by the lower panel of Fig. \ref{fig:zahid12}.
Therefore, we also show the \citet{zahid12} model with $R$ and $P_{\rm O}$ chosen to reproduce the simulated $M_{\rm O,tot}$, and $\Delta Z$ chosen to reproduce either $M_{\rm O,*}$ (blue dashed line) or $M_{\rm O,gas}$ (green dot-dashed line).
There is no combination of model parameters that can simultaneously reproduce the simulated trends of $M_{\rm O,*}$, $M_{\rm O,gas}$, and $M_{\rm O,tot}$.
This may be because their model uses observed trends for gas-phase metallicity, which may be subject to systematic uncertainties in absolute metallicity calibrations \citep{kewley08,kewley19}.
More fundamentally, regardless of the parameter values used, the shape of $1-\fo$ as a function of $M_*$ predicted by the \citet{zahid12} model is the same as in Fig. \ref{fig:comp}, i.e. increasing $1-\fo$ with increasing $M_*$.
This may indicate that the analytic functions used in the model are insufficient to properly represent the necessary physical processes, however a more detailed investigation is beyond the scope of this work.}

Our cosmological simulations can reproduce many observations, including the mass--metallicity relations, $M_{\rm BH}$--$\sigma$ relation \citet{pt15a}, and galaxy metallicity gradients \citep{pt16,pt17b}.
There are several possible reasons for the discrepancies seen between our model and other works.
%It may also be the case that our models do not properly capture the physical processes necessary to reproduce the observations.
If observational estimates are much higher than our predictions, then it could require updates to our modelling of feedback and metal ejection: in our current models, feedback from both supernovae and AGN is isotropic and purely thermal, and such a prescription has proved successful for reproducing the properties of galaxies seen on $\sim$ kpc scales.
However feedback processes that take place on significantly smaller scales, such as the formation of large-scale SNe-driven winds in highly star-forming galaxies, or the launching of an AGN jet, may lead to pathways out of a galaxy along which metals can more easily escape \citep[e.g.,][]{fielding18}.

There is not yet a consistent definition in the literature of where metals can be considered lost from galaxies.
The virial radius might seem an obvious choice so that temporarily ejected material is included, but measuring the metallicity of the circumgalactic medium can only be done along quasar sight lines \citep[e.g.,][]{tumlinson11}.
From within the virial radius, enriched gas may still fall back onto the galaxy {as a fountain} \citep[e.g.,][]{tremblay18}, and it is not clear whether these metals should be thought of as lost, or still part of the galaxy system.
\citet{ck07} considered all gas particles that had ever been present in a galaxy, regardless of whether they were in the galaxy at the present day.
Hydrodynamic simulations do not suffer from this observational limitation; using the FIRE simulations, \citet{ma16} measured the fraction of metals retained within a virial radius of each galaxy, finding consistent values with this work (see Fig. \ref{fig:R200SFMS}).

The radial dependence of metal loss must be considered, rather than an integrated quantity, since this can more readily and robustly be compared between observations and simulations.
\citet{belfiore16} analysed NGC 628 in such a way, finding that metal loss was highest in the bulge ($\sim 70$ per cent lost) and decreased towards the outskirts, with about half of all oxygen lost out to 7 kpc.
This is the opposite trend to Fig. \ref{fig:maps}, but we note that the stellar mass of NGC 628 is estimated to be $10^{9.8}\msun$, and so may be more susceptible to losing material than the comparatively high mass galaxies of Fig. \ref{fig:maps}.
In a future paper (Taylor et al., in prep.) we will investigate this radial dependence of metal loss for our simulated galaxies.

\section{Conclusions}\label{sec:discussion}

We have predicted the fraction of oxygen mass lost from galaxies of different masses and in different environments.
The amount of oxygen that is lost depends both on feedback from SNe and AGN as well as the chemical evolution model, and can therefore be a strong test of galaxy evolution models.
Presently, very few observational measurements have been made to enable detailed comparison with models; observations have been mostly restricted to nearby, isolated dwarf galaxies for which individual stars can be resolved \citep[e.g.,][]{kirby11,mcquinn15}.

In this work we have, for the first time, investigated the effect of environment on the fraction of metals lost from galaxies.
In addition to reproducing the trend of lower mass galaxies losing more metals seen in both observational and theoretical works \citep{dalcanton07,ma16,du17}, we make several predictions, which are summarised below:
\begin{itemize}
	\item Galaxies that make up the star formation main sequence also occupy a tight sequence in terms of metal retention -- the metal flow main sequence, ZFMS -- (Fig. \ref{fig:foms}).
	\item Quenched, massive galaxies tend to have lost a greater fraction of metals than star-forming galaxies at the same mass, due to the ejection of metals via AGN-driven winds to the CGM and IGM \citep[see also][]{pt15b}.
	\item Satellite galaxies in richer groups tend to have lost a greater fraction of their oxygen following gas loss due to environmental quenching (Fig. \ref{fig:env}).
	\item Most isolated galaxies (those with no companions within $R_{200}$) lie on the ZFMS, but stellar or AGN feedback (depending on the galaxy mass) can cause them to lose more metals.
	\item On galaxy scales, we find that the majority of star-forming galaxies show an enhancement in oxygen compared to the amount that could have been produced by their population of stars, whereas most quenched galaxies have a net loss of oxygen (Fig. \ref{fig:foms}). On halo scales, oxygen is conserved or lost across the entire galaxy population (Fig. \ref{fig:R200SFMS}), suggesting a net inflow of metals into the centres of star-forming galaxies.
\end{itemize}

%%%%%%%%%%%%%% conclusions %%%%%%%%%%%%%%%%%
%\section{Conclusions}
%\label{sec:conc}

\section*{Acknowledgements}

We thank the referee, Jorge Moreno, for his useful comments, which improved the quality of this paper.
Parts of this research were supported by the Australian Research Council Centre of Excellence for All Sky Astrophysics in 3 Dimensions (ASTRO3D), through project number CE170100013.
Parts of this manuscript were written at an ASTRO3D-funded writing retreat.
CK acknowledges funding from the UK Science and Technology Facility Council (STFC) through grants ST/M000958/1 and ST/R000905/1.
This work has made use of the University of Hertfordshire Science and Technology Research Institute high-performance computing facility.
This work used the DiRAC Data Centric system at Durham University, operated by the Institute for Computational Cosmology on behalf of the STFC DiRAC HPC Facility (www.dirac.ac.uk). This equipment was funded by BIS National E-infrastructure capital grant ST/K00042X/1, STFC capital grants ST/H008519/1 and ST/K00087X/1, STFC DiRAC Operations grant  ST/K003267/1 and Durham University. DiRAC is part of the National E-Infrastructure.
PT thanks A.~Medling, T.~Mendel, and B.~Groves for helpful discussions.
This work was born out of conversations held at the workshop ``Metals in Galaxies, Near and Far: Looking Ahead", hosted by the Lorentz Centre, Leiden, {following a presentation and discussion by K.~McQuinn}.
Finally, we thank V.~Springel for providing {\sc GADGET-3}.

\section*{Data Availability}
The data underlying this article will be shared on reasonable request to the corresponding author.

%%%%%%%%%%%%%%  bibliography  %%%%%%%%%%%%%%%%%

\bibliographystyle{mn2e}
\bibliography{/Users/ptaylor/papers/refs}
%\begin{thebibliography}{}

%\end{thebibliography}

%%%%%%%%%%%%%%  appendix  %%%%%%%%%%%%%%%%%
%\appendix

\bsp

\label{lastpage}

\end{document}